# THE VERIFICATION OF VIRTUAL COMMUNITY MEMBER'S SOCIO-DEMOGRAPHIC PROFILE

Fedushko Solomia[1], Peleschyshyn Oksana[2], Peleschyshyn Andriy[1], Syerov Yuriy[1]

[1]Social Communications and Information Activities Department, Lviv Polytechnic National University, Ukraine
felomia@gmail.com

[2]Mathematical Modelling of Socio-Economic Processes Department, Ivan Franko National University of Lviv, Ukraine

## ABSTRACT

*This article considers the current problem of investigation and development of method of web-members' socio-demographic characteristics' profile validation based on analysis of socio-demographic characteristics. The topicality of the paper is determined by the necessity to identify the web-community member by means of computer-linguistic analysis of their information track (all information about web-community members, which posted on the Internet). The formal model of basic socio-demographic characteristics of virtual communities' member is formed. The algorithm of these characteristics verification is developed.*

## 1. INTRODUCTION

Web-communities [1,6] accumulated a huge database of contacts and profiles, which contain a lot of information about the person. The auditorium of web-communities is a large number of people regardless of age, gender, occupation, education, ethnicity, social status etc., who in order to register need to fill in a form with their personal data [4]. Nowadays, the necessity of verifying of web-community members' personal data is the topical issue.

The value of this research lies in verifying basic socio-demographic characteristics (SDCh) of communities' member based on comprehensive analysis of information track of community members which gives an opportunity to the identifying socio-demographic profile (SD profile) of virtual community's member [5]. Every socio-demographic characteristic of virtual community's member is determining by analysis of linguistic features in virtual community's members communication. The difference in style of writing posts by virtual community's members is the basis for developing effective methods of verifying of personal data in users account.

These issues have the greatest influence on efficiency rise of virtual communities functioning and the level of data authenticity [8] in Web-members' personal profiles. Solution to these problems is possible by using computer-linguistic analysis of web-members' posts.

The actual problem of modern web is "information noise" [3] – a huge amount of inexact, incomplete and superficial information, which Internet users created on the Web.

To achieve the aim, the following research tasks should be fulfilled: to develop the algorithm for authenticity verification of the socio-demographic characteristics of web-community's members and model of the basic socio-demographic characteristics of communities' member.

Thus, computer-linguistic analysis of basic socio-demographic characteristics can be of essential help to administrators and moderators to monitor and validate personal data provided by users of the web-community. Also, this analysis improves the techniques of effective targeted advertising campaign.

## 2. THE VERIFICATION OF VIRTUAL COMMUNITY MEMBER'S SOCIO-DEMOGRAPHIC PROFILE

### 2.1. Definition of SDCh in different fields of science

Analysis of socio-demographic characteristics of virtual community's member lies in the forming socio-demographic profile's of web-community's member. The notion "socio-demographic characteristics" [5] is necessary to consider because these characteristics are base for formation of SDP of web-community's member. In social communication the SDCh are defined as a set of basic characteristics of the account in web community.

The socio-demographic characteristics are provided by member of this web-community. However, the authenticity of these data is always questionable.

In general, the socio-demographic characteristics include: age, gender, material security, position, marital status, education level, profession and work experience, employment, location, religious and political views, etc.

Thus, the socio-demographic characteristics are a set of social estimation criteria and important parameters of human activity.

The notion of socio-demographic characteristics has a wide range of use in many sciences. Socio-demographic characteristics are defined in various fields as important parameters.
The results of thorough analysis of using social-demographic characteristics in scientific investigations are shown on Fig.1

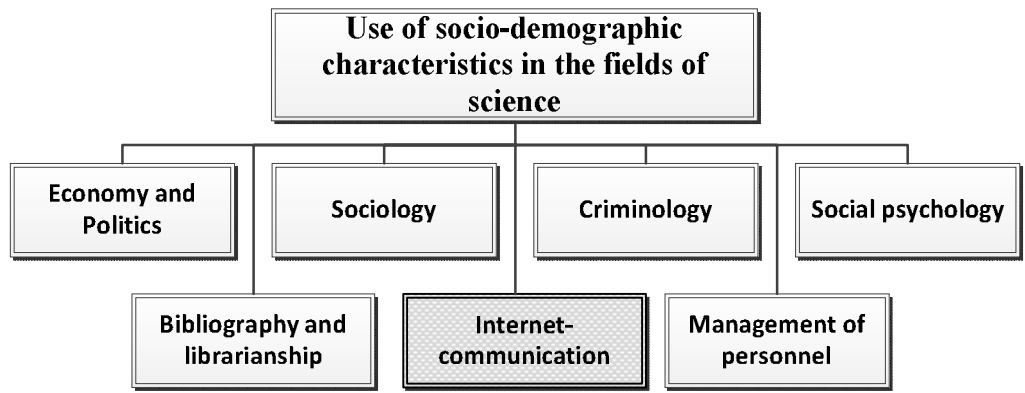

Fig.1 Use of SDCh in the fields of science

Thus, the socio-demographic identity of community member various social communications of global network study many fields of sciences [5]: sociology, psychology, culture, criminology, management and economics, etc.

However, the fundamental application of this work will be implemented in the internet-marketing. The authenticity of socio-demographic characteristics in the process of using online advertisement has a important value.

## 2.2 Use of the SDCh in the internet marketing

The popularity of virtual communities among Internet users [7,8] causes the feasibility of using communication capabilities of social media in various fields of human activity (by private users and organizations).

Construction process of internet communications includes: determination of virtual communities goal, choice of specific types of communities and methods of interaction. Adaptation of the behaviours, methods of information dissemination and regulations of social media is an important factor of influence on the effectiveness of communications and the attainability of goal. Identifying socio-demographic characteristics of virtual communities' members and taking these characteristics into account in the communication process organization are essential.

The use of virtual communities in the marketing activities of the modern company has a positive impact on company image and expands opportunities to promote the company.

The topical areas of internet marketing of a company are:

– establishing the company representation in virtual communities and its positioning;
– organization of customer support system and community of clients based on the website of the company;
– Internet advertising of the company, its website and products.

The realization of these areas is associated with virtual communities.

Nowadays almost every company that carries on actively marketing activities, has its own official website and other related websites (for branding, social events, etc). The basic socio-demographic characteristics of the target audience (the actual and potential consumers of products and services of the company) in the design development and content websites creation are certainly taking into consideration.

The main tasks for company management, as a system of representations in virtual communities:

– establishment and regular updating of marketing materials in the form of posts and discussions in communities;
– tracking and responding to the development of discussions;
– recording and monitoring of representative materials in social media.

The discussion style, content and form of information materials, which are located on the web pages of communities, must comply with the general character of the communication process in every social media. This is achieved because the company specialists in planning and implementation of marketing activities in a virtual community considered the socio-demographic characteristics of the participants.

The customers' community organization around company is an effective creation method of the consumers set. Consumers actively promote the company and its products, share experiences, helping to improve the products and often provide a high level of mutual support (especially in cases of high-tech products) and form the knowledge base for effective exploitation of company products. The indication of the high level of confidence to the company and the popularity of its products is the presence of an organized consumer's community.

The implementation of customer support through virtual communities is realized by searching and monitoring thoughts and communities members' feedback on the company products. Customers are posting their requests and applications in discussions. The responses to requests and applications are often posted in the same discussion. The ability to actively search and identify the needs and requests of web users, which previously did not handed over for processing by the company, is essential. This approach significantly expands the company potential regarding communication with users. Nevertheless, the presented approach is sufficiently laborious.

The concept of leadership in the ideas field is often used to promote the sale and promotion of new products, services and market research organization. Opinion leaders [2] – target buyers in the market, who often give advice related to a category of products or services. Determining the values of certain socio-demographic characteristics (such as level of education, professional experience, etc.), which are peculiar features of the majority opinion leaders, is effective in selection of virtual communities' members, who influence on opinions of the others consumer audience.

Thus, the employment of consumers groups in the popular Internet social networking to organization the online-communication of company "friends" is a topical issue. During the work with consumers in social networks have to be taken into account age and other specialized social networks. During the work with consumers in social networks the age and other specialization (age, gender and others) of social networks. Specialization of social networking is displayed on the theme and style of discussion in thematic groups.

## 2.3 A formal model of basic socio-demographic characteristics of virtual communities' member

We introduce a formal model of the basic socio-demographic characteristics (SDCh) of virtual communities' member. Our SDCh model expresses the basic socio-demographic characteristics as ordered set of reliable socio-demographic characteristics of virtual community's member, and is described by a mathematical equation. The socio-demographic characteristics model includes only basic socio-demographic characteristics of virtual communities' member, such as "age", "sphere of activity", "gender" and "level of education". These socio-demographic characteristics take into consideration, because it is useful for problem of internet-marketing.

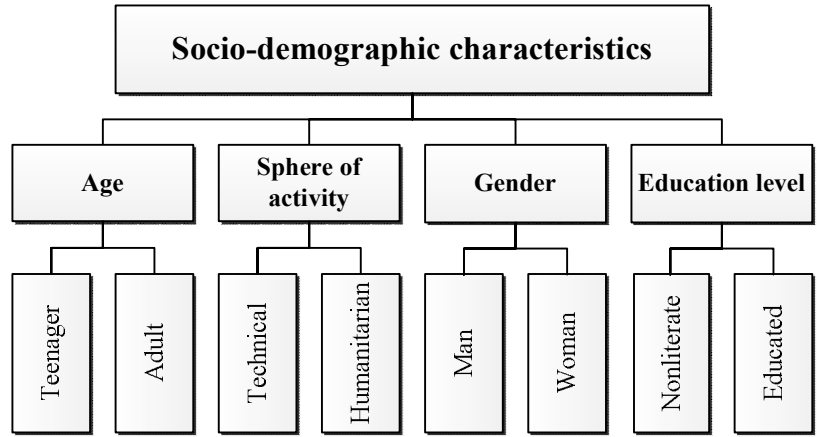

Fig.2 The structure of the investigated SDCh of virtual community member

The linguistic and communication indicators determine the community member belonging to a particular set of socio-demographic characteristics.

Linguistic and communication indicators - special feature of language and communication of online community's member, which can be traced in his information track.

Researchers defined the information track as a set of all personal data of virtual community's member and the results of his communicative activity - the content, which is created by web member. The socio-demographic characteristics model of member virtual community – socio-demographic profile – describe as follows:

$$SDP(U^*) = \left(SDCh_j(U^*)\right)_{j=1}^{N^{SDCh(U^*)}}, \quad (1)$$

where $\left(SDCh_j(U^*)\right)_{j=1}^{N^{SDCh(U^*)}}$ – ordered set of socio-demographic characteristics of community member $U^*$; $N^{SDCh}(U^*)$ – quantity of these characteristics of member $U^*$.

In the particular case, a set of socio-demographic characteristics (see Fig.2) is described as:

$$SDCh(U^*) = \left(age(U^*), edu(U^*), gend(U^*), sphere(U^*)\right), \quad (2)$$

The *"age"-SDCh* of virtual community's member *age(U\*)* takes one of two values: age(U*) ∈ {"teenager"; "adult"}.

In conformity with mental human development at age stages in developmental psychology [4], web-community members into two groups are distributed: "teenager" (from 6 to 17 years old) and "adult" (18+ years old).

The *"age"-SDCh* value of virtual community's member *age (U \*)* is determined by the vector of age indicators *Age*:

$$Age = \left(Age_j(U^*)\right)_{j=1}^{N^{Age}}, \quad (3)$$

where $Age_j(U^*)$ – j-age-indicator of member U*, $N^{Age}$ – quantity of age indicators.
The *"education level"-SDCh* of virtual community's member $edu(U^*)$ takes one of two values: $edu(U^*) \in \{"educated"; "nonliterate"\}$.

The distribution of *"education level"-SDCh* on the value realized on the basis of web content, that created by the member of web-community. It allows to judge of the education level of community member.

The *"education level"-SDCh* value of web member $edu(U^*)$ is determined by the vector of education indicators *Edu*:

$$Edu = \left(Edu_j\left(U^*\right)\right)_{j=1}^{N^{Edu}} \quad , \quad (4)$$

where $Edu_j(U^*)$ – j-indicator of educational level, $N^{Edu}$ – quantity of educational level indicators.
The *"gender"-SDCh* of virtual community's member $gend(U^*)$ takes one of two values: $edu(U^*) \in \{"man"; "woman"\}$.

Gender differentiation of virtual communities' members defines "gender"-SDCh of community members. The separation of the community members occurs according to biological characteristic: "man" and "woman".
The *"gender"-SDCh* value of web member $gend(U^*)$ is determined by the vector of gender indicators *Gend*:

$$Gend = \left(Gend_j\left(U^*\right)\right)_{j=1}^{N^{Gend}} \quad , \quad (5)$$

where $Gend_j(U^*)$ – j-gender-indicator, $N^{Gend}$ – quantity of gender-indicators.

The *"sphere of activity"-SDCh* of virtual community's member $sphere(U^*)$ takes one of two values: $sphere(U^*) \in \{"technical"; "humanitarian"\}$.

This division of community member sphere of activity is effective in determination the probable profession, hobbies and fields of activities of community's member, namely, the sphere in which the person is qualified.

The *"sphere of activity"-SDCh* value of web member $sphere(U^*)$ is determined by the vector of indicators of sphere of activity *Sphere*:

$$Sphere = \left(Sphere_j\left(U^*\right)\right)_{j=1}^{N^{Sphere}} \quad , \quad (6)$$

where $Sphere_j(U^*)$ – j-indicator of sphere of activity of web-community's member U*, $N^{Sphere}$ – quantity of sphere of activity indicators.

According to formal model of basic socio-demographic characteristics of virtual communities' member, web-community's member has characterized four socio-demographic characteristics. Each socio-demographic characteristics of virtual community's member takes two values, which describe the community member. These data are the basis for developing an algorithm of the verification of socio-demographic characteristics of virtual community's member.

## 2.4 The algorithm for verification of socio-demographic characteristics

Analysis of information track (a set of all personal data of virtual community member and the results of his communicative activity) of community members which take place according to algorithm of validation of account data on web-community member allows to detect authenticity level of data in web-members' personal profiles.

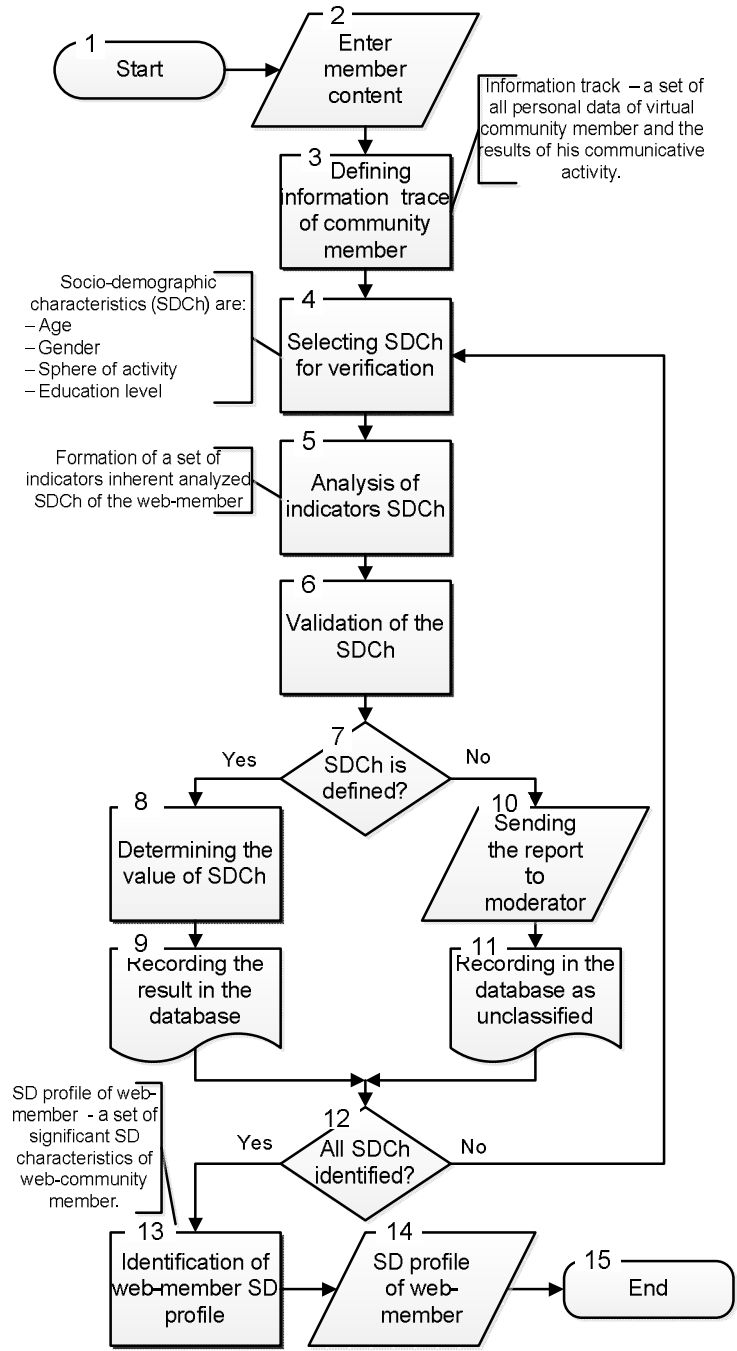

Fig. 3 Block diagram of verification of SD Profile algorithm

Socio-demographic profile of virtual community's member [5] – a set of significant socio-demographic characteristics of web-community member.

Algorithm of verification of socio-demographic characteristics of web-community's members is presented on the Figure 3.

The idea of the algorithm contains the verifying and information analysis of the members' data. This algorithm is formed on the basis of specialized software which has been implemented in work of "Lviv.Ridne Misto Web-forum" (see Fig. 4).

## 2.5 The practical results of verification of SD Profile algorithm

The result of the algorithm is classification of Web-forum members' accounts which can helps an administrator to arrange Web-forum members' accounts, promptly monitored the personal data in Web-forum members' accounts and socio-demographic profile of virtual community's member.

Fig. 4 Socio-demographic profiles of members "Lviv.Ridne Misto Web-forum"

The main aim of this work is to create the method of computer-linguistic analysis of web-communities members' information track.

Analysis of data verifying of members which will occur according to algorithm of verification of socio-demographic characteristics of web-community's members allows to detected level of authenticity of data in web-members' personal profiles and to separated members with reliable and unreliable data.

Classification of web-forum members' accounts into level of data reliability:

- account with reliable information;
- account data under suspicion;
- account with inadequate information (pseudo user account).

Advanced Computing: An International Journal ( ACIJ ), Vol.4, No.3, May 2013*Account with reliable information.* Personal information which gives member in his account is reliable or member changed it at the request of web-community administration which is sent during computer-linguistic analysis. The communicative behaviour of the member conforms to all web-community regulations.

*Account data under suspicion.* Not all personal data of web-members are reliable. The member partially or completely ignored the request to change the incorrect information. All data changes and communicative behaviour are monitored by web-community administration.

*Account with inadequate information* (Pseudo user account). The data in personal profile is completely unauthentic. The web-forum member contravenes the web-community regulations. The member profile is blocked / deleted or access to the web community is banned.

## 3. CONCLUSION

A question of urgent importance in the web-forum management and moderation is the development of a new approach to data verification which gives community members when they are resisted.

The method of personal data validation of the maximum amount of information of web-community member is developed and basic foundation for computer-linguistic method of web-forum members' information track verification, which is based on linguistic analysis of web-community content, is devised.

The issue of socio-demographic profile of virtual community's member verifying is investigated. The socio-demographic profile of virtual community's member helps automatically administrated and monitored the "Lviv.Ridne Misto Web-forum".

Thus, this paper presents a new approach to developing computer-linguistic method of community member identification by socio-demographic characteristics of web-community's members and identifying community members by means of computer-linguistic analysis of web-communities members' information tracks.

**Authors**

**Solomia Fedushko**

Solomia is PhD student and assistant at the Social Communications and Information Activities Department of the Lviv Polytechnic National University in Ukraine. She holds a Master's degree in Applied Linguistics. Her current research interests are in internet linguistics, social communications in Web, web-communities, web-community content analysis. She has published numerous scientific articles.

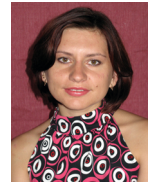

**Oksana Peleschyshyn**

Oksana is a PhD student at Department of Mathematical modelling of socio-economic processes, Ivan Franko National University of Lviv in Ukraine. Her research interests include marketing in online communities.

**Andriy Peleschyshyn**

Andriy is Professor and Head of Social Communications and Information Activities Department of the Lviv Polytechnic National University. He holds a Ph.D. from Lviv Polytechnic National University. He has published numerous articles and books in the areas of Global Information technology, social network position, web-mining technologies, Social Communications and Information Activities.

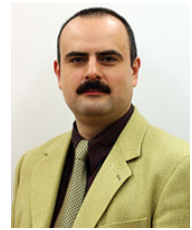

**Yuriy Syerov**

Yuriy holds a master's degree in computer science or computer engineering, Lviv Polytechnic National University. He holds a Ph.D. in technical sciences from Lviv Polytechnic National University in Ukraine, in 2010. At the present moment: Associated Professor at the Social Communications and Information Activities Department. Scientific interests: web-communities, organization of web-communities, web-mining technologies.

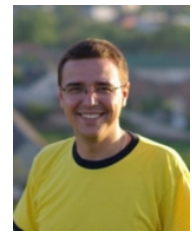